\documentclass[10pt,aps,prc]{revtex4}

\usepackage{epsfig}
\usepackage{dcolumn} 
\usepackage{amsfonts}
\usepackage{color}

\newcommand{\be}{\begin{equation}}
\newcommand{\ee}{\end{equation}}
\newcommand{\ba}{\begin{array}}
\newcommand{\ea}{\end{array}}
\newcommand{\bea}{\begin{eqnarray}}
\newcommand{\eea}{\end{eqnarray}}

\def\half{{1\over2}}

\def\Rb{\mathbb{R}}
\def\Cb{\mathbb{C}}

\begin{document}


\title{An algebraic approach to problems with polynomial Hamiltonians on Euclidean
spaces}


\author{D.J.~Rowe}
\affiliation{Department of Physics, University of Toronto\\
Toronto, Ontario M5S 1A7, Canada}

\begin{abstract} 
Explicit expressions are given for the actions and radial matrix
elements of basic radial observables on multi-dimensional spaces in a continuous
sequence of orthonormal bases for unitary SU(1,1) irreps.
Explicit expressions are also given for  SO$(N)$-reduced matrix elements of basic
orbital observables. These developments make it possible to determine the matrix
elements of polynomial and a other Hamiltonians analytically, to within SO$(N)$
Clebsch-Gordan coefficients, and to select an optimal basis for a particular problem
such that the expansion of eigenfunctions is most rapidly convergent.  
\end{abstract}
\maketitle 

\section{Introduction} 

This paper is concerned with the large class of problems lying between the
relatively few that are exactly solvable and others that are only solvable
by numerical methods. 
In particular, algebraic methods are developed for computing the spectral properties
of Hamiltonians which are polynomials in the Cartesian position $\{ x_i\}$ and
momentum
$\{ \hat p_i = -i \hbar \partial/\partial x_i\}$ observables of a real Euclidean
space $\Rb^N$; polynomials in the inverse square of the radial coordinate are also
considered.
Restriction to this sub-class of Hamiltonians is because, for them it is
appropriate to employ bases of harmonic oscillator or modified oscillator wave
functions.
Probably a similar treatment can be developed for hydrogenic and modified
hyrdrogenic bases which will enable the admission of potentials, like the   
$1/r$ potential of the hydrogen atom, which are not rational functions of the
Cartesian coordinates.

The most widely studied Hamiltonians of this kind are for exactly
solvable central force problems.
Some central force problems are solved by use of various realizations of an SU(1,1)
spectrum generating algebra (cf., for example, Refs.\ \cite{BGW,CP,CW} for reviews).
Others are solved \cite{others} by the so-called {\em factorization method\/}
introduced by Schr\"odinger \cite{Schr} and employed extensively by Infeld and Hull
\cite{IH}. (A pedagogical review of the factorization method is given in several
texts on quantum mechanics, e.g.\ \cite{HechtQM}.) 
Many special case solutions for power law and inverse power law potentials have also
been obtained by the methods of Refs.\ \cite{Singh,Znojil,Bose,Kar}.
In an attempt to identify the underlying symmetries of exactly solvable
systems, Gendenshtein \cite{Genden} introduced the criterion of {\em shape
invariance\/} and a connection with supersymmetry.
Exactly solvable central force problems play an essential role in
the current investigation.
However, their purpose will be to provide suitable bases in terms
of which the matrix elements of a much larger class of Hamiltonians of interest can
be determined algebraically.

Underlying the solvability of a central force problem is the factorization of its
Hilbert space into a product of radial and orbital subspaces 
\be \mathcal{L}^2(\Rb^N) \simeq \mathcal{L}^2(\Rb^+,dr) \otimes
\mathcal{L}^2(S_{N-1}) ,
\ee
where $\Rb_+$ is the positive half of the real line,
\be  \mathcal{L}^2(\Rb^+, dr) = \left\{ R: r \to \Cb \ \Big| 
\int_0^\infty \big|R(r)\big|^2\, dr <\infty \right\}, \label{eq:L2R+}\ee
and the elements of $\mathcal{L}^2(S_{N-1})$ are square integrable functions on the
$N-1$ sphere, $S_{N-1}\simeq {\rm SO}(N)/{\rm SO}(N-1)$, relative to the
SO$(N)$--invariant measure. 
Thus, because the Hamiltonian for a central force problem is SO$(N)$--invariant its
Schr\"odinger equation  reduces to a one-dimensional radial equation.

More generally the radial $\Rb^+$ and orbital $S_{N-1}$ dynamics are coupled by
centrifugal forces.
Nevertheless, the ${\rm SU}(1,1)\times {\rm O}(N)$ dynamical group associated with
the above factorization continues to provide powerful algebraic (and hence
exact) methods for determining the matrix elements of polynomial Hamiltonians in
suitable bases.  This is of enormous practical value even if it is ultimately
necessary to resort to numerical methods for diagonalizing the Hamiltonian matrix.
Moreover, it is a huge advantage if bases
functions can be found which are close to the desired solutions so that the
dimensions of the matrices that have to be diagonalized to achieve a given level of
accuracy are relatively small.

With such application in mind, Armstrong \cite{Arm}, Haskell and Wybourne \cite{HW},
for example, derived expressions for some radial matrix elements in a basis of
eigenstates of the three-dimensional harmonic oscillator.
It turns out that, for many purposes, other matrix elements are needed and 
are also needed for higher-dimensional spaces.
 For example, the nuclear collective model is defined on a five-dimensional space
\cite{BM,Rowe,RT}.
And, while harmonic oscillator bases are appropriate for spherical vibrational
systems, they are not the most appropriate for rotational-vibrational systems such
as diatomic molecules and non-spherical nuclei.

 The current paper gives analytical expressions for the
matrix elements of needed observables in a continuous sequence of
orthonormal bases for unitary ${\rm SU}(1,1)\times {\rm O}(N)$ irreps. 
These irreps include, but are not
restricted to, the irreps of the harmonic (discrete) series. 
Nor are they restricted to a three-dimensional space.
Matrix elements of observables that lie in the SU(1,1) Lie algebra 
are obtained by algebraic methods.
Factorization methods are used to compute
the actions and matrix elements of other needed observables.
 These developments make it possible to select an optimal basis for a particular
problem and continue to retain analytical expressions for matrix elements as provided
by the harmonic oscillator bases.
 Basis functions for  $\mathcal{L}^2(S_{N-1})$, which span O$(N)$
irreps, are commonly referred to as SO$(N)$-spherical harmonics.
They are known for SO(3), ${\rm SO}(4)\simeq {\rm SU}(2)\times{\rm SU}(2)$, SO(5)
\cite{CMS,RTR}, and SO(6). 
However, as the following shows, explicit expressions for the wave functions are not
needed for the computation of matrix elements.
What more is needed, beyond the expressions given in this paper, 
are the appropriate Clebsch-Gordan coefficients.
These are now available in SO(3)-coupled bases for all $N\leq 6$  (cf.\ section
\ref{sect:concl}).

\section{Representations of ${\rm SU(1,1)}\times{\rm SO}(N)$}
  
The SU(1,1) Lie algebra (more precisely its complex extension) is spanned by
operators $\{\hat S_\pm,\hat S_0\}$ which satisfy the standard commutation relations
\be [\hat S_-, \hat S_+] = 2 \hat S_0 \,, \quad 
[\hat S_0,  \hat S_\pm] = \pm\hat S_\pm \,.
\ee
A basis for an SU(1,1) irrep is given by an orthonormal set of states 
$\{ |\lambda \nu\rangle; \nu = 0,1,2, \dots\}$ which satisfy the identities
\be \begin{array}{ccc} 
\hat S_0 |\lambda \nu\rangle = \half
(\lambda+2\nu)|\lambda \nu\rangle\,,\\ 
\phantom{\Big|} \hat S_+ |\lambda \nu\rangle = \sqrt{(\lambda +\nu)(\nu+1)}\,
|\lambda ,\nu+1\rangle\,, \\ 
 \hat S_- |\lambda \nu\rangle = \sqrt{(\lambda +\nu-1)\nu}\, |\lambda
,\nu-1\rangle\,. 
\end{array}  \label{eq:S0+-}
\ee
For such an irrep the Casimir invariant
\be \hat C_{\rm SU11} = (\hat S_0)^2 - \half (\hat S_+ \hat S_- + 
\hat S_- \hat S_+) = \hat S_0(\hat S_0-1) -\hat S_+ \hat S_- \ee
takes the value $\lambda (\lambda -2)/4$, i.e.,
\be \hat C_{\rm SU11} |\lambda\nu\rangle = \frac{1}{4} \lambda (\lambda -2)|\lambda
\nu\rangle .
\ee

The SO$(N)$ Lie algebra is spanned by angular momentum operators $\{
\hat\mathcal{L}_{ij}\}$ which are antisymmetric, $ \hat\mathcal{L}_{ij} = -
\hat\mathcal{L}_{ji}$, and satisfy the commutation relations
\be \big[ \hat\mathcal{L}_{ij},\hat\mathcal{L}_{kl}\big] = -{\rm i}\left[\delta_{jk}
\hat\mathcal{L}_{il} - \delta_{il} \hat\mathcal{L}_{kj}\right].
\ee
The SO$(N)$ Casimir operator is defined by
\be \hat \Lambda = \sum_{i<j} \hat\mathcal{L}_{ij}^2 . \label{eq:SONCas}
\ee

If an SO$(N)$ irrep, labelled by $v$, has basis states $\{ |vm\rangle \}$, then
a basis for an
${\rm SU(1,1)}\times{\rm SO}(N)$ irrep $(\lambda,v)$ is given by the product states
\be |\lambda\nu; vm\rangle \equiv |\lambda \nu\rangle \times |vm\rangle
.\label{eq:basisstates}\ee

\subsection{Harmonic series representations}

An explicit realization of the SU(1,1) Lie algebra is given in terms of  harmonic
oscillator raising and lowering operators
\be c^\dag_i = {1\over \sqrt{2}} \Big( \hat x_i -
\frac{\partial}{\partial x_i}\Big) \,, \quad
 c_i = {1\over \sqrt{2}} \Big( \hat x_i +
\frac{\partial}{\partial x_i}\Big) \,, \quad
i = 1, \dots ,N,\label{eq:cops}\ee
by the O$(N)$-invariant operators
\be \begin{array}{cccc} \hat S_+ = \displaystyle\half  c^\dag\cdot
c^\dag\,,\quad
\hat S_- = \half c\cdot c \,, \\
\hat S_0 = \displaystyle {1\over 4} ( c^\dag\cdot c +c\cdot c^\dag )
= \half\left( c^\dag\cdot c + \frac{N}{2}\right)
=\half\left(\hat n + \frac{N}{2}\right), \label{eq:Sops}
\end{array}
\ee
where $c\cdot c = \sum_i c_ic_i$ and $\hat n =\sum_i c^\dag_i c_i$ is a harmonic
oscillator number operator.
Similarly the SO$(N)$ angular momentum operators take the form
\be {\cal L}_{ij} = -{\rm i} ( c^\dag_i c_j - c^\dag_j c_i) \,.
\ee

When the SU(1,1) and SO$(N)$ operators are realized in this way, 
their Casimir invariants are given, respectively, by
\be 4\hat C_{\rm SU11} =  \hat n (\hat n+N-2) +\frac{1}{4}
N(N-4) - (c^\dag \cdot c^\dag) (c\cdot c)\ee
and
\be \hat \Lambda = \hat n (\hat n+N-2)- (c^\dag \cdot c^\dag) (c\cdot c) .\ee
It is seen that  when acting  on the Hilbert space
$\mathcal{L}^2(\Rb^N)$, according to the above-defined
realizations, these Casimir invariants are linearly related by
the identity
\be \hat\Lambda = 4\hat C_{\rm SU11} - \frac{1}{4} N(N-4) ,\ee
in accordance with a well-known duality relationship  \cite{Howe} (also called
complementarity \cite{MQ}). This duality implies that the labels
$\lambda$ and $v$ are in one-to-one correspondence.
Thus, if the state $|\lambda 0;v\sigma\rangle $ satisfies  
the equations
\be \hat S_- |\lambda 0; v\sigma\rangle = 0, \quad \hat n |\lambda 0; v\sigma\rangle
= v|\lambda 0; v\sigma\rangle ,
\ee
then $\lambda$ takes the value
\be  \lambda = v+ N/2 , \quad v= 0,\; 1,\; 2,\; \cdots \label{eq:HOlambdav}\ee
Moreover, the eigenvalue of $\hat\Lambda$ 
\be \hat\Lambda |\lambda 0; vm\rangle = v(v+N-2) |\lambda 0; vm\rangle
\ee
signifies that $v$, often referred as {\em seniority\/}, is an SO$(N)$ angular
momentum quantum number.

\subsection{Modified oscillator representations}

The discrete harmonic series of SU(1,1) irreps which have duality relationships with
SO$(N)$ irreps are invaluable for many purposes.  
However, there is a continuous
series of irreps, which are often more useful.
These more general realization of the SU(1,1) algebra are obtained
by expanding the above operators, using eqns.\ (\ref{eq:cops}),
\be  
\hat S_\pm = \displaystyle {1\over 4} \left[ \nabla^2 + r^2
\mp \left(2 r\frac{\partial}{\partial r} +N\right)\right]  , \quad
  \hat S_0 = \displaystyle{1\over 4} \left[-\nabla^2  +  r^2 \right] , \label{eq:19}
\ee
where $r^2 = \sum_i x_i^2$, and  $\nabla^2 = \sum_i \partial^2/\partial x_i^2$ is the
Laplacian on
$\mathcal{L}^2(\Rb^N)$.  The Laplacian has the well-known expansion
\be
\nabla^2 =  \hat\Delta - \hat{\Lambda\over r^2}
\,,\label{eq:nabla2}
\ee
where
\be \hat\Delta = {1\over r^{N-1}} {\partial\over\partial r} r^{N-1}
{\partial\over\partial r} .
\ee

In this form the SU(1,1) operators act as differential operators on the wave
functions for the basis states $\{ |\lambda \nu ; vm\rangle\}$ of equation
(\ref{eq:basisstates}). These wave functions are conveniently expanded
\be \Psi_{\lambda\nu vm}({\bf r}) = r^{-(N-1)/2}\, \mathcal{R}^\lambda_\nu
(r)\, \mathcal{Y}_{vm}(\omega) , \label{eq:wfns}\ee
where, for convenience, the factor of $r^{N-1}$ in the $\Rb^N$ volume element
 $dv(r,\omega) = r^{N-1}\, dr\, d\omega$ is absorbed into the definition of
the wave functions. 
Thus,  $\{ \mathcal{R}^\lambda_\nu ; \nu = 0,1,2,\dots\}$ is a basis of
radial wave function for $\mathcal{L}^2(\Rb^+, dr)$ and
$\{\mathcal{Y}_{vm}(\omega)\}$ is a basis of spherical harmonics for
$\mathcal{L}^2(S_{N-1})$ defined as eigenfunctions of $\hat\Lambda$
\be \hat\Lambda \mathcal{Y}_{vm} = v(v+N-2)\,\mathcal{Y}_{vm}.
\ee

With the Laplacian in the form $\nabla^2 = \hat\Delta -\hat\Lambda /r^2$,
the action of the SU(1,1) operators of equation (\ref{eq:Sops}) on the wave functions of
eqn.\ (\ref{eq:wfns}) gives
\bea &\left[\hat S_\pm \Psi_{\lambda\nu vm}\right]\!( r,\omega) = 
 r^{-(N-1)/2} \left[\hat \mathcal{S}_\pm^{(\lambda)} \,
\mathcal{R}^\lambda_\nu\right]\!( r)\, \mathcal{Y}_{vm}(\omega) , & \label{eq:24}\\
& \left[\hat S_0 \Psi_{\lambda\nu vm}\right]\!( r,\theta) = 
 r^{-(N-1)/2} \left[\hat\mathcal{S}_0^{(\lambda)} \,
\mathcal{R}^\lambda_\nu\right]\!( r)\, \mathcal{Y}_{vm}(\omega) , &\label{eq:25}
\eea
with $\lambda = v + {N/2}$ and
\bea  
&\hat\mathcal{S}_\pm^{(\lambda)} = \displaystyle {1\over 4} \left[
\frac{d^2}{dr^2}  -\frac{(\lambda- 3/2)(\lambda-1/2)}{ r^2} +
 r^2
\mp \left(2 r\frac{d}{dr} +1\right)\right]  , & \label{eq:Spmlambda}\\
& \hat\mathcal{S}_0^{(\lambda)}= \displaystyle{1\over 4} \left[-
\frac{d^2}{dr^2}  +\frac{(\lambda- 3/2)(\lambda-1/2)}{ r^2} +
 r^2 \right]. & \label{eq:S0lambda}
\eea

Although the $\hat S^{(\lambda)}$ operators have been derived
for the discrete values of $\lambda = v+N/2$, they nevertheless satisfy
the SU(1,1) commutation relations
\be [\hat\mathcal{S}_-^{(\lambda)}, \hat\mathcal{S}_+^{(\lambda)}] = 
2 \hat\mathcal{S}_0^{(\lambda)}
\,, \quad  [\hat\mathcal{S}_0^{(\lambda)},\hat\mathcal{S}_\pm^{(\lambda)}] =
\hat\mathcal{S}_\pm^{(\lambda)} ,
\ee
for any value of $\lambda$.
Moreover, for any real $\lambda > 0$, they have a unitary representation and define 
an orthonormal basis $\{\mathcal{R}^{\lambda}_n\}$ 
for the Hilbert space $\mathcal{L}^2(\Rb_+)$ such that
\be \begin{array}{ccc} 
\hat \mathcal{S}^{(\lambda)}_0 {\cal R}^\lambda_{\nu} = \half
(\lambda+2\nu){\cal R}^\lambda_{\nu}\,,\\ 
\phantom{\Big|} \hat \mathcal{S}^{(\lambda)}_+{\cal R}^\lambda_{\nu} = \sqrt{(\lambda
+\nu)(\nu+1)}\, {\cal R}^\lambda_{\nu+1}\,, \\ 
 \hat \mathcal{S}^{(\lambda)}_- {\cal R}^\lambda_{\nu} = \sqrt{(\lambda +\nu-1)\nu}\,
{\cal R}^\lambda_{\nu-1} \,. 
\end{array}  \label{eq:radeqns}
\ee
For each $\lambda>0$, the $\hat S^{(\lambda)}$ operators span a specific
SU(1,1) representation. This is evidenced by the fact that, for each realization, 
the SU(1,1) Casimir operator acquires the unique value
$\lambda (\lambda-2)/4$, i.e.,
\be \hat C_{\rm SU11}^{(\lambda)} = \hat\mathcal{S}_0^{(\lambda)}(
\hat\mathcal{S}_0^{(\lambda)} -1) -
\hat\mathcal{S}_+^{(\lambda)}\hat\mathcal{S}_-^{(\lambda)}  = \frac{1}{4}\lambda
(\lambda-2) \hat I,
\ee
where $\hat I$ is the identity.

\section{The radial wave functions}

\begin{figure} 
\epsfxsize=2.9in {\epsfbox{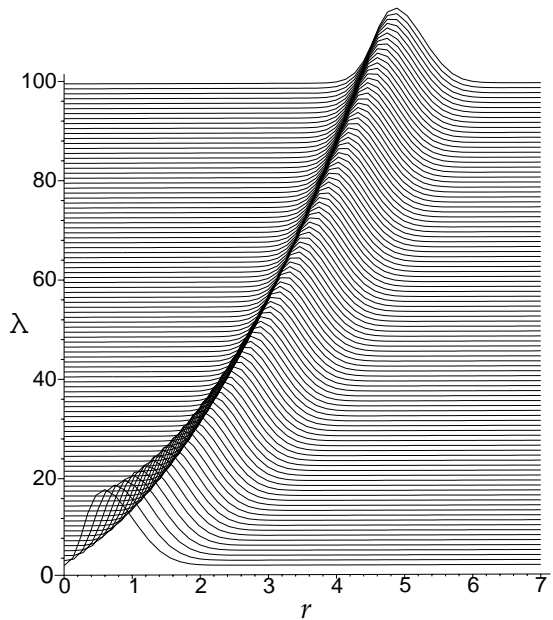}}
  \caption{Radial wave functions $\{ \mathcal{R}^{\lambda}_0\}$ plotted as
   functions of $r$ for a range of values of $\lambda$. (Computed
by P.S.\ Turner \cite{RT}.)\label{fig:DavWFS}} 
\end{figure}

The  $\{\mathcal{R}^{\lambda}_\nu\}$ wave functions for arbitrary $\lambda > 0$
are derived as for the standard harmonic oscillator and are given by 
\be {\mathcal R}^{\lambda}_\nu(r)
 = (-1)^\nu\sqrt{{2\nu!\over\Gamma(\lambda +\nu)}} \: 
r^{\lambda-1/2}    \: {\rm L}_\nu^{(\lambda-1)}(r^2)
\: e^{-r^2 /2} , \quad \nu = 0,\; 1,\; 2,\; \dots\label{eq:Rfns}
\ee 
The lowest-weight functions $\{\mathcal{R}^\lambda_0\}$ are plotted in Fig.\ 
\ref{fig:DavWFS}.

An important characteristic of the radial Hilbert space $\mathcal{L}^2(\Rb^+,dr)$ is
that it is independent of the dimensionality of the Euclidean space in which it is
embedded.  
Thus, when $2\lambda$ is an integer, these functions  are equal to the radial wave
functions for various $N$-dimensional harmonic oscillators.
For example, the wave functions of the simple harmonic oscillator are  given by
\bea &\displaystyle u_{2n}(x) = \frac{1}{\sqrt{2}}\mathcal{R}^{1/2}_n (x) = 
(-1)^n \frac{2^{n}\, n!}{\sqrt{\sqrt{\pi} (2n)!}}\, L^{(-1/2)}_n(x^2)\,
e^{-x^2/2} ,& \quad \infty < x< \infty ,\\
&\displaystyle u_{2n+1}(x) = \frac{1}{\sqrt{2}}\mathcal{R}^{3/2}_n (x) = 
(-1)^n \frac{2^{n+1/2}\, n!}{\sqrt{\sqrt{\pi} (2n+1)!}}\, xL^{(1/2)}_n(x^2)\,
e^{-x^2/2} ,&\quad \infty < x< \infty .
\eea
Radial wave functions for the spherical
$(N=3)$ harmonic oscillator are  given by
\be u_{nl}(r)  = {\mathcal R}^{l+3/2}_n(r)=
(-1)^n \sqrt{{2n!\over\Gamma(n+l+3/2)}}
\:  r^{l+1} \,{\rm L}_n^{(l+1/2)}(r^2)\, e^{-r^2 /2} . 
\ee
In general, the radial wave functions for an $N$-dimensional
harmonic oscillator of SO$(N)$ angular momentum $v$ are given by
\be u_{nv}(r) =\mathcal{R}^{v+N/2}_n(r).\ee
These identities reflect the fact that  harmonic oscillator radial wave functions
belong to SU(1,1) irreps which are determined by the angular momentum of the
accompanying spherical wave function. For example, eigenstates of the
three-dimensional harmonic oscillator of angular momentum $l$ have wave functions
which in spherical polar coordinates take the form
\be \Psi_{nlm}(r,\theta,\varphi) = \frac{1}{r} u_{nl}(r)\, Y_{lm}(\theta,\varphi),\ee
where $Y_{lm}$ is a spherical harmonic.

When $2\lambda$ is not an integer, the basis functions $\{\mathcal{R}^\lambda_n\}$
are not the radial wave functions of any harmonic oscillator.
Nevertheless, they have algebraic 
properties that are almost as simple as those which are. 
For many systems, especially models that exhibit rotational as
well as vibrational states, it is appropriate to use  
$\{\mathcal{R}^{\lambda}_n\}$ radial wave functions with $\lambda > v+N/2$.
Such wave functions are eigenfunctions of a modified oscillator Hamiltonian
introduced by Davidson \cite{Dav}  for the purpose of describing the
rotational-vibrational states of a diatomic molecule. 
In this model, the atoms of the molecule are `pushed apart' 
by enhancing the centrifugal potential by the substitution
\be
 \frac{l(l+1)}{r^2} \;\; \to \;\; \frac{l(l+1)+r_0^4}{r^2} ;
\ee
this is equivalent to replacing the harmonic oscillator potential $r^2$ by a
potential
\be   V(r)=\frac{r_0^4}{r^2} + r^2 \ee
which has a minimum at $r=r_0$.
A five-dimensional version of the Davidson oscillator has been considered as a
model for nuclear rotations and vibrations \cite{Roh,EEP,RB98}.

\section{Radial matrix elements}

\subsection{Matrix elements obtained from irreps of the SU(1,1) Lie
algebra}\label{sect:SU11}

{\bf Claim 1:} For $\lambda >1$,
\bea r^2{\cal R}^\lambda_\nu (r)&\!\!=\!\!&
\sqrt{(\lambda +\nu-1)\nu}\,   
{\cal R}_{\nu-1}^\lambda(r) +\sqrt{(\lambda + \nu)(\nu+1)}\,
{\cal R}_{\nu+1}^\lambda(r) \nonumber\\
&& \quad + (\lambda+2\nu)\,{\cal R}_\nu^\lambda(r)   ,
\label{eq:claimI1}\\
\frac{1}{r^2} \,{\cal R}^\lambda_\nu(r) &\!\!=\!\!& 
\sum_{\mu <\nu} \frac{(-1)^{\mu -\nu}}{\lambda-1} \sqrt{\frac{\nu!\,
\Gamma(\lambda+\mu)}{\mu!\,
\Gamma(\lambda+\nu)}}\, {\cal R}^\lambda_\mu(r) 
\nonumber\\
&&+\sum_{\mu \geq\nu}
\frac{(-1)^{\mu -\nu}}{\lambda-1} \sqrt{\frac{\mu!\,
\Gamma(\lambda+\nu)}{\nu!\,
\Gamma(\lambda+\mu)}}\, {\cal R}^\lambda_\mu (r),\label{eq:claimI2}\\ 
\frac{d^2}{dr^2} {\cal R}^\lambda_{\nu}(r)
&\!\!=\!\!& \sqrt{(\lambda +\nu-1)\nu}\, {\cal R}^{\lambda }_{\nu-1}(r)
+\sqrt{(\lambda +\nu)(\nu+1)}\, {\cal R}^{\lambda}_{\nu+1}(r) 
\nonumber \\ 
&& -  (\lambda +2\nu) {\cal R}^{\lambda}_{\nu}(r) 
 + (\lambda-3/2)(\lambda- 1/2) \frac{1}{r^2} 
{\cal R}^\lambda_{\nu}(r) \,. \label{eq:claimI3} \\
\nabla^2 \mathcal{R}^\lambda_\nu(r) \mathcal{Y}_{vm}(\omega)
&\!\!=\!\!&  \mathcal{Y}_{vm}(\omega) 
\left[ \hat S^{(\lambda)}_+ +\hat S^{(\lambda)}_- -2\hat S^{(\lambda)}_0 
+ \frac{(\lambda-1)^2 - (v+ \half N-1)^2}{r^2} \right]\mathcal{R}^\lambda_\nu(r).
\label{eq:claimI4}
\eea
\medskip

The first of these equations follows from equation (\ref{eq:radeqns}) and 
the observation that
$r^2$ is an element of the su(1,1) Lie algebra with expansion
\be
 r^2 = \hat S^{(\lambda)}_+ + \hat S^{(\lambda)}_- + 2\hat S^{(\lambda)}_0 \,.
\ee
Equation (\ref{eq:claimI2}) is obtained by rexpressing  equation
(\ref{eq:claimI1}) as the recursion relation
\be 
(\lambda+2\nu)\, f^\lambda_{\mu\nu} + \sqrt{(\lambda + \nu-1)\nu}\,
f^\lambda_{\mu,\nu-1} +\sqrt{(\lambda + \nu)(\nu+1)}\, f^\lambda_{\mu,\nu+1} =
\delta_{\mu\nu} \,, \label{eq:326}
\ee
where
\be 
f^\lambda_{\mu\nu}= \int {\cal R}^\lambda_\mu(r)\,
\frac{1}{r^2}\, {\cal R}^\lambda_\nu(r) 
\, {\rm d}r =f^\lambda_{\nu\mu} \,.
\ee
Starting with the value of $f^\lambda_{00} = 1/(\lambda-1)$ obtained from
equation (\ref{eq:145}), this equation has solution
\be 
f^\lambda_{\mu\nu} = \frac{(-1)^{\mu-\nu}}{\lambda-1} \sqrt{\frac{\nu!\, 
\Gamma(\lambda+\mu)}{\mu!\,
\Gamma(\lambda+\nu)}}\,, \quad {\rm for}\; \mu\leq \nu\,.
\ee
Equation (\ref{eq:claimI3}) follows from the identity
\be  \frac{d^2}{dr^2} -
\frac{(\lambda-\frac{3}{2})(\lambda-\frac{1}{2})}{r^2}
= \hat S^{(\lambda)}_+ + \hat S^{(\lambda)}_- - 2\hat S^{(\lambda)}_0 \,.
\ee
The expression for $\nabla^2$ is obtained by recalling that, from
equations (\ref{eq:19}), (\ref{eq:24})
and (\ref{eq:25}),
\bea &\displaystyle\langle \lambda\mu;
v\big|\nabla^2\big|\lambda
\nu;v\rangle = 
\langle\lambda\mu \big|\Big[ \hat S^{(v+\half N)}_+ + \hat
S^{(v+\half N)}_-  - 2\hat S^{(v+\half N)}_0\Big]
\big|\lambda \nu\rangle \qquad\qquad\qquad&\nonumber\\
&= \displaystyle\langle\lambda\mu \big|\Big[ \hat S^{(\lambda)}_+ + \hat
S^{(\lambda)}_-  - 2\hat S^{(\lambda)}_0 +
\frac{(\lambda\!-\!\half)(\lambda\!-\!\frac{3}{2})
-(v\!+\!\half N \!-\!\half)(v\!+\!\half N \!-\!\frac{3}{2})}{r^2}\Big] 
\big|\lambda \nu\rangle \,. &
\eea
\medskip

Matrix elements of higher even powers of $r$ are given by repeated use of equation
(\ref{eq:claimI1}).

\subsection{Matrix elements obtained by the factorization method}
\label{sect:factorization} 

Similar equations for the operators $r$, $1/r$, and
$d/dr$, are obtained  by  the factorization  method \cite{IH,HechtQM}).
Let $A(X)$ and $A^\dag(X)$, where $X$ is a real number, denote the operators
\be 
A(X)=\frac{d}{dr}+\frac{X}{r} + r, \quad  A^\dag(X) = -\frac{d}{dr}+\frac{X}{r} + r
 .\ee Then
\bea A(X) A^\dag (X) = -\frac{d^2}{dr^2} + \frac{X(X-1)}{r^2} + r^2+ 2X+1 , \\
A(X)^\dag  A(X) = -\frac{d^2}{dr^2} + \frac{X(X+1)}{r^2}  + r^2+ 2X-1 .
\eea
It follows that
\bea  
&A(\lambda-\frac{1}{2}) A^\dag (\lambda-\frac{1}{2}) = 4 \mathcal{S}_0^{(\lambda)} +
2\lambda,& \label{eq:47}\\  
&A^\dag(\lambda-\frac{1}{2}) A (\lambda-\frac{1}{2}) = 4
\mathcal{S}_0^{(\lambda+1)} + 2\lambda -2,& \label{eq:48}\\
&A(-\lambda+\frac{3}{2}) A^\dag (-\lambda+\frac{3}{2}) = 4 \mathcal{S}_0^{(\lambda)}
- 2\lambda +4,& \label{eq:49}\\  
&A^\dag(-\lambda+\frac{3}{2}) A (-\lambda+\frac{3}{2}) = 4
\mathcal{S}_0^{(\lambda-1)} - 2\lambda +2.&\label{eq:50}
\eea

\begin{figure} 
\epsfxsize=5in {\epsfbox{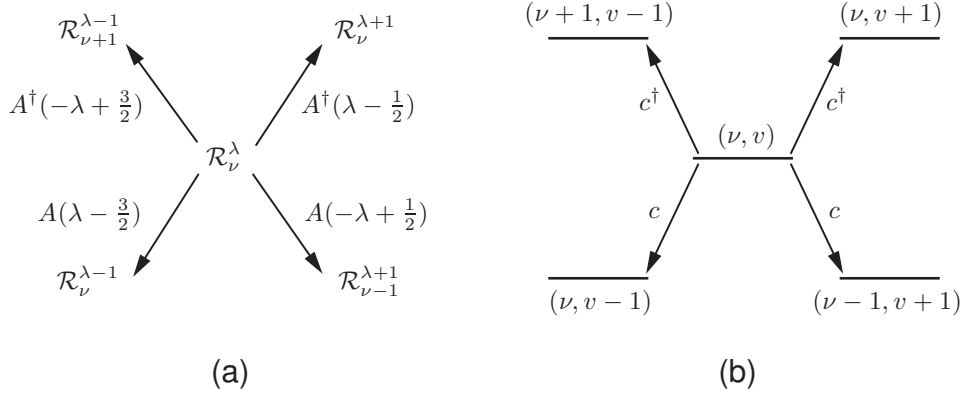}}
  \caption{The action of the raising and lowering operators of radial wave
functions compared to the actions of the standard raising and lowering 
operators of harmonic oscillator states. 
Multiplets of harmonic oscillator states of a given energy level are shown as
horizontal lines in (b) labelled by radial
$\nu$ and SO$(N)$ angular momentum $v$.
\label{fig2a}}
\end{figure}

Equations (\ref{eq:47}) and (\ref{eq:48}) give
\bea  
& A(\lambda-\frac{1}{2}) A^\dag (\lambda-\frac{1}{2}) \mathcal{R}^{\lambda}_\nu = 
4(\lambda +\nu)\mathcal{R}^{\lambda}_\nu  ,&\label{eq:51}\\  
&A^\dag(\lambda-\frac{1}{2}) A (\lambda-\frac{1}{2}) \mathcal{R}^{\lambda+1}_\nu = 
4(\lambda +\nu)\mathcal{R}^{\lambda+1}_\nu ,& \label{eq:52}
\eea
and hence
\bea  
&A^\dag(\lambda-\frac{1}{2}) A (\lambda-\frac{1}{2}) 
\left[A^\dag(\lambda-\frac{1}{2})\mathcal{R}^{\lambda}_\nu\right] = 
4(\lambda +\nu)\left[A^\dag(\lambda-\frac{1}{2})\mathcal{R}^{\lambda}_\nu\right],&
\label{eq:53}\\ 
& A(\lambda-\frac{1}{2}) A^\dag (\lambda-\frac{1}{2})
\left[A(\lambda-\frac{1}{2})\mathcal{R}^{\lambda+1}_\nu\right] =  4(\lambda+\nu)
\left[A(\lambda-\frac{1}{2})\mathcal{R}^{\lambda+1}_\nu\right].&
\label{eq:54}  
\eea
From these and similar equations obtained from equations (\ref{eq:49}) and
(\ref{eq:50}), it follows (with a choice of relative phase) that
\bea & A^\dag(\lambda-\frac{1}{2})\mathcal{R}^{\lambda}_\nu = 
2\sqrt{\lambda+\nu}\ \mathcal{R}^{\lambda+1}_\nu ,&\label{eq:67}\\
& A(\lambda-\frac{1}{2})\mathcal{R}^{\lambda+1}_\nu =
2\sqrt{\lambda +\nu}\ \mathcal{R}^{\lambda}_{\nu} , &\label{eq:68}\\
 & A^\dag(-\lambda+\frac{3}{2})\mathcal{R}^{\lambda}_\nu = 
2\sqrt{\nu+1}\ \mathcal{R}^{\lambda-1}_{\nu+1} ,
\quad {\rm for}\;\;\lambda >1& \label{eq:69}\\
 & A(-\lambda+\frac{3}{2})\mathcal{R}^{\lambda-1}_{\nu+1} =
2\sqrt{\nu+1}\ \mathcal{R}^{\lambda}_{\nu},
\quad {\rm for}\;\;\lambda >1 . \label{eq:70}
\eea
Thus, it is seen that the $A$ and $A^\dag$ operator act as raising and lowering
operators for the radial wave functions in close parallel with the way the standard
raising and lowering operators act on harmonic oscillator states; the parallel is
exhibited in fig.\ \ref{fig2a}.
This observation is exploited in Section \ref{sect:combinedMEs} to obtain the
actions of the harmonic oscillator raising and lowering operators on general modified
oscillator states.

From the identities (\ref{eq:67}--\ref{eq:70}), we obtain the results: 

{\bf Claim 2:} For $\lambda >1$
\bea 
r \,{\cal R}^\lambda_\nu(r) &\!\!=\!\!& 
\sqrt{\lambda +\nu-1}\,{\cal R}^{\lambda-1}_{\nu}(r)
+ \sqrt{\nu+1}\, \,{\cal R}^{\lambda-1}_{\nu+1}(r) ,\label{eq:claimII1}\\
r \,{\cal R}^{\lambda}_\nu(r) &\!\!=\!\!& 
 \sqrt{\lambda +\nu}\,{\cal R}^{\lambda+1}_\nu(r)
+ \sqrt{\nu}\, \,{\cal R}^{\lambda+1}_{\nu-1}(r) , \label{eq:claimII2}\\
\frac{1}{r} {\cal R}^\lambda_\nu(r) &\!\!=\!\!& \sum_{\mu =0}^\nu
(-1)^{\mu-\nu}
\sqrt{\frac{\nu!\,\Gamma(\lambda +\mu-1)}{\mu!\,\Gamma(\lambda +\nu)}}\,
{\cal R}^{\lambda -1}_\mu (r), \qquad \label{eq:claimII3}\\
\frac{1}{r} {\cal R}^\lambda_\nu(r) &\!\!=\!\!& \sum_{\mu =\nu}^\infty
(-1)^{\mu-\nu}
\sqrt{\frac{\mu!\,\Gamma(\lambda +\nu)}{\nu!\,\Gamma(\lambda +\mu+1)}}\,
{\cal R}^{\lambda +1}_\mu (r),\label{eq:claimII4}\\
\frac{d}{dr} {\cal R}^\lambda_\nu(r) 
&\!\!=\!\!&  -\sqrt{\nu+1}\, {\cal R}^{\lambda-1}_{\nu+1}(r) 
+ \frac{\nu+\half}{\sqrt{\lambda +\nu-1}}{\cal R}^{\lambda-1}_{\nu}(r)
\nonumber\\ 
&&- (\lambda -\frac{3}{2})\sum_{\mu =0}^{\nu-1} (-1)^{\mu-\nu}
\sqrt{\frac{\nu!\,\Gamma(\lambda +\mu -1)}{\mu!\,\Gamma(\lambda +\nu)}}\,
{\cal R}^{\lambda -1}_\mu (r), \quad\label{eq:claimII5} \\
\frac{d}{dr} {\cal R}^\lambda_\nu(r) &\!\!=\!\!& 
\sqrt{\nu}\, {\cal R}^{\lambda +1}_{\nu-1}(r) 
- \frac{\nu +\half}{\sqrt{\lambda +\nu}}{\cal R}^{\lambda+1}_{\nu}(r)
\nonumber\\ 
&&+ (\lambda -\half)\sum_{\mu =\nu+1}^{\infty} (-1)^{\mu-\nu}
\sqrt{\frac{\mu!\,\Gamma(\lambda +\nu)}{\nu!\,\Gamma(\lambda +\mu +1)}}\,
{\cal R}^{\lambda +1}_\mu (r). \label{eq:claimII6} 
\eea

The first of these equations follows from equations (\ref{eq:68}) and
(\ref{eq:69}) and the observation that
\be
 2r = A^\dag (-\lambda +\textstyle\frac{3}{2}) + A
   (\lambda -\frac{3}{2})\,. \label{eq:2r1}
\ee
The second follows similarly from equations (\ref{eq:67}) and 
(\ref{eq:70}) with
\be
 2r = A^\dag (\lambda-\textstyle\half) + A (-\lambda+\half)\,. \label{eq:2r2}
\ee
Equation (\ref{eq:claimII3})  is obtained by expressing equation
(\ref{eq:claimII2}) as a recursion relation
\be \frac{1}{r} {\cal R}^\lambda_\nu(r) = \frac{1}{\sqrt{\lambda
+\nu-1}}\, {\mathcal R}^{\lambda-1}_\nu(r) -
\sqrt{\frac{\nu}{\lambda+\nu-1}}\, \frac{1}{r} {\mathcal
R}^\lambda_{\nu-1}(r) ,
\label{eq:145}\ee
which is readily solved to give the desired result.
Equation (\ref{eq:claimII4})  is similarly obtained from equation
(\ref{eq:claimII1}).
The last two equations of the claim are obtained from the identity
\be A(X) -A^\dag(-X) = {2}\left(\frac{d}{dr} +
\frac{X}{r}\right).
\ee
Setting $X= \lambda -3/2$, and using equation (\ref{eq:claimII3})
leads to equation (\ref{eq:claimII5}).
Setting $X= -\lambda +1/2$, and using equation (\ref{eq:claimII4})
leads to equation (\ref{eq:claimII6}).

\subsection{An O$(N)$-parity quantum number} \label{sect:parity}

The above results show that radial matrix elements of type
$\langle \lambda \mu|r^2|\lambda \nu\rangle $, 
$\langle \lambda \mu|1/r^2|\lambda \nu\rangle $, and
$\langle \lambda \mu|d^2/dr^2|\lambda \nu\rangle$ are obtained by use of the SU(1,1)
Lie algebra and those of type
$\langle \lambda \pm 1, \mu|r|\lambda \nu\rangle $, 
$\langle \lambda \pm 1, \mu|1/r|\lambda \nu\rangle $, and
$\langle \lambda \pm 1, \mu|d/dr|\lambda \nu\rangle$ are obtained by use of the
factorization method.
However, neither of the methods presented gives  matrix elements of the type
$\langle \lambda \mu|r|\lambda \nu\rangle $, 
$\langle \lambda  \mu|1/r|\lambda \nu\rangle $, and
$\langle \lambda \mu|d/dr|\lambda \nu\rangle$.
This is because odd powers of $r$ and $d/dr$ do not occur alone in the space of
polynomial functions of the basic $\{ x_i,p_i\}$ observables.
For example, whereas $r^2$ is the quadratic $r^2=\sum_i x_i^2$, there is no
polynomial expression for $r =\sqrt{\sum_i x_i^2}$.
However, terms linear in $r$ and $d/dr$ do occur in combination
with orbital functions.   For example, when expressed in terms of $\Rb^N$
spherical polar coordinates, the Euclidean coordinates $\{ x_i\}$ are of the form
$x_i = r \mathcal{Q}_i (\omega)$, where $\mathcal{Q}_{i}$ is proportional to a $v=1$
SO$(N)$ spherical harmonic.

The implications of this observation are conveniently summarized in terms of an
O$(N)$-parity quantum number
$\pi = (-1)^v$  associated with every SO$(N)$ irrep of angular momentum $v$.
It is seen that even and odd functions of the $\{ x_i\}$ coordinates have
even and odd O$(N)$-parity, respectively.
It follows that the matrix elements 
$\langle \lambda' \mu; v'm |\hat x_i |\lambda\nu ; vn\rangle$ and
$\langle \lambda' \mu; v'm |\hat p_i |\lambda\nu ; vn\rangle$
vanish unless $(-1)^{v'} = (-1)^{v+1}$.
This means that, in the evaluation of polynomial observables, the matrix elements of
$r$ and $d/dr$, for example, are only needed between states of opposite parity.
Thus, the results of Sections \ref{sect:SU11} and
\ref{sect:factorization} lead to algebraic expressions for matrix
elements of (positive and negative) integer powers  of $r$ and $d/dr$ if one chooses
basis states $\{ |\lambda\nu;vm\rangle\}$ with $\lambda$ related to $v$ by
\be \lambda_{v+1} = \lambda_v  \pm 1.\ee
The standard relationship $\lambda_v = v+N/2$, given by equation (\ref{eq:HOlambdav})
for the harmonic series of SU(1,1) irreps, obviously satisfies this condition.
However, there are many other possibilities including, for example, having just two
SU(1,1) irreps, one for even- and one for odd O$(N)$-parity states such that
\be \lambda_+ = \lambda_- \pm 1 \,.\ee

It should be ephasized that 
Hamiltonians that are not polynomials in $\{ x_i,p_i\}$ are also of interest.
The Hamiltonian of the hydrogen atom, with a singular $1/r$ potential, is a
standard example. However, it would appear that, for such Hamiltonians, the radial
wave functions of the modified oscillator do not provide the most appropropriate
basis wave functions.

\section{Orbital matrix elements}

This section gives SO$(N)$-reduced matrix elements of the basic $v=1$, SO$(N)$
tensor, $\mathcal{Q}$, defined by the expression
\be x_i = r \mathcal{Q}_i
\ee
 of the Cartesian  coordinates in SO$(N)$ spherical polar coordinates.
To make use of these reduced matrix elements one will, in general, need 
access to SO$(N)$ Clebsch-Gordan coefficients.
The required CG ocefficients are known in SO(3) $\supset$ SO(2) coupled bases for
$N\leq 6$ (cf.\ Concluding remarks).

\subsection{SO$(N)$-reduced matrix elements and a symmetry property}

SO$(N)$-reduced matrix elements are defined by the Wigner-Eckart theorem
\be \langle v_3m_3 |\mathcal{Q}_{m_2} | v_1m_1\rangle = (v_1m_1,\! 1m_2 |v_3m_3)\,
\langle v_3 \| \mathcal{Q}\| v_1\rangle ,
\ee
where $(v_1m_1,\! 1m_2 |v_3m_3)$ is an SO$(N)$ Clebsch-Gordan coefficient and
$|vm\rangle$ is an orbital basis state whose wave function is an SO$(N)$ spherical
harmonic  $\mathcal{Y}_{vm}$.
It is then useful to define the SO$(N)$-coupled action of a tensor operator, e.g.\
$\mathcal{Q}$, by 
\bea \left[ \mathcal{Q}\otimes |v_1\rangle \right]_{v_3m_3} &=& \sum_{m_1m_2}
|v_3m_3\rangle \langle v_3m_3 |\mathcal{Q}_{m_2} |v_1m_1\rangle\,\! (v_1m_1, 1m_2
|v_3m_3) \nonumber\\
&=& |v_3m_3\rangle \langle v_3 \| \mathcal{Q}\| v_1\rangle .
\eea

An application of these definitions leads to the useful expression
\be 
\int \left[\mathcal{Y}_{v_3}(\omega) \otimes\mathcal{Q}
\otimes\mathcal{Y}_{v_1}(\omega)\right]_0\, d\omega 
= k_{v_3} \langle v_3 \| \mathcal{Q}\| v_1\rangle  ,
\ee
and, hence, the identity
\be k_{v_3} \langle v_3 \| \mathcal{Q}\| v_1\rangle = k_{v_1} \langle v_1 \|
\mathcal{Q}\| v_3\rangle ,
\ee
where
\be k_v =  \int \left[ \mathcal{Y}_{v}(\omega)
\otimes\mathcal{Y}_{v}(\omega)\right]_0 \, d\omega .\ee
The value of $k_v$ can be inferred, to within a phase factor, by making the expansion
\be \mathcal{Y}_{vn} = \sum_n C_{nm} \mathcal{Y}^*_{vm} ,\ee
which  is always possible because the space $\mathcal{L}^2(S_N)$ has a real basis.
Because spherical harmonics are defined to be an orthonormal basis of orbital wave
functions, it follow that
\be \sum_{n} |C_{nm}|^2 = 1 , \quad  \ee
and that
\be k_v = \sum_{m'mn} C_{nm'} (vm,vn |00) 
\int \mathcal{Y}^*_{vm'}(\omega)\,\mathcal{Y}_{vm}(\omega) \, d\omega 
=\sum_{mn} C_{nm} (vm,vn |00) .
\ee
Now, from the two identities
\be \sum_{mn} |C_{nm}|^2 = d(v) , \quad \sum_{mn}(vm,vn |00)^2 = 1, \ee
where $d(v)$ is the dimension of the SO$(N)$ irrep $v$, we conclude that
\be C_{nm} = k_v (vm,vn |00) \quad \mathrm{and} \quad |k_v|^2 = d(v) .\ee
It follows that
\be k_v = e^{\mathrm{i}\phi(v)}
\sqrt{d(v)} ,
\ee
where $\phi(v)$ is a phase angle, and 
\be \langle v_3 \| \mathcal{Q}\| v_1\rangle = e^{\mathrm{i} (\phi(v_1)-\phi(v_3))}
\sqrt\frac{d(v_1)}{d(v_3)}\ \langle v_1 \|\mathcal{Q}\| v_3\rangle \,.
\label{eq:RedMRsymmetry}
\ee

{It is important to note, that reduced matrix elements are sometimes defined by
expressing the Wigner-Eckart theorem in the form
$$\langle v_3m_3 |\mathcal{Q}_{m_2} | v_1m_1\rangle = (v_1m_1,\! 1m_2 |v_3m_3)\,
\frac{\overline{\langle v_3 \| \mathcal{Q}\| v_1\rangle}}{\sqrt{d(v_3)}} .$$
This adjusted definition of the reduced matrix elements results in a simplification
of the symmetry relation (\ref{eq:RedMRsymmetry}) to 
\be \overline{\langle v_3 \| \mathcal{Q}\| v_1\rangle} = e^{\mathrm{i}
(\phi(v_1)-\phi(v_3))}\,
\overline{\langle v_1 \|\mathcal{Q}\| v_3\rangle} \,.
\ee}

\subsection{Basic reduced  matrix elements}

Reduced matrix elements of the basic $v=1$, SO$(N)$ tensor, $\mathcal{Q}$, are
determined by the expansion
\be x_i = \frac{1}{\sqrt{2}} ( c^\dag_i + c_i )\ee
and the matrix elements of the harmonic oscillator raising and lowering
operators.
This expression shows that $\mathcal{Q}$ has non-zero reduced matrix
elements $\langle v'\| \mathcal{Q}\|v\rangle$ only for $v'=v\pm 1$.

Let 
\be a^\dag =\frac{1}{\sqrt{2}} ( c_1^\dag + \mathrm{i} c_2^\dag) \ee
denote the highest weight component of the harmonic oscillator raising operators 
in which the operators $\{ c^\dag_i\}$ of equation (\ref{eq:cops}) are regarded as
Cartesian components of a $v=1$, SO$(N)$ tensor.
Then,  with $\lambda_v = v+N/2$, the harmonic oscillator ground state is the state
$|0\rangle = |\lambda_0 \nu=0; v=m=0\rangle$ and a subset of excited states is given
by
\be  |\lambda_v 0; vv\rangle = \frac{1}{\sqrt{v!}} \left( a^\dag\right)^v
|0\rangle, \quad v= 0,\; 1,\; 2,\; \cdots\ee
The matrix element 
\be \langle \lambda_{v+1}0; v+1,v+1| a^\dag |\lambda_v 0; vv\rangle
= \sqrt{v+1}
\ee
then implies that
\be \langle\lambda_{v+1}0; v+1\| r\mathcal{Q} \|\lambda_v 0; v\rangle =
\frac{1}{\sqrt{2}}\langle \lambda_{v+1}0; v+1\| c^\dag
\|\lambda_v 0; v\rangle = \sqrt{\frac{v+1}{2}} . \label{eq:rQme}\ee
Thus, we obtain the following result:

\medskip{\bf Claim 3:} SO$(N)$-reduced matrix elements of the basic $v=1$, SO$(N)$
tensor, $\mathcal{Q}$, defined by the expression $x_i = r \mathcal{Q}_i$ are given
by
\be  \langle v'\| \mathcal{Q} \| v\rangle =\sqrt{\frac{v+1}{2v+N}}\ \delta_{v',v+1}
+   e^{\mathrm{i} (\phi(v)-\phi(v-1))}\
\sqrt\frac{d(v)\,v}{d(v-1) (2v+N-2)}\ \delta_{v',v-1} .
\ee
\medskip

The first term of this expression is obtained by factoring out the  radial matrix
element 
 $\langle\lambda_v+1,0 |r|\lambda_v 0\rangle = \sqrt{\lambda_v} =\sqrt{v+N/2}$
in equation (\ref{eq:rQme}).
The second term is obtained from the first by use of the symmetry relationship 
(\ref{eq:RedMRsymmetry}).

For $N=3$, for example, the customary expression of the Cartesian coordinates in
terms of $(r,\theta,\varphi)$ spherical polar coordinates 
\bea &x_1 = r\, \sin\theta\, \cos\varphi ,&\nonumber\\
    & x_2 = r\, \sin\theta\, \sin\varphi,& \label{eq:Q1}\\
    & x_3 = r\, \cos\theta , &\nonumber
\eea
defines the Cartesian components $\{ \mathcal{Q}_i\}$  of the $N=3$
tensor $\mathcal{Q}$. 
The dimension of an SO(3) irrep of angular momentum $l$
(a standard symbol for SO(3) angular momentum) is given by
$d(l) = 2l+1$ and, with the standard phase convention for SO(3) spherical harmonics, 
we obtain $e^{\mathrm{i}\phi(l)} = (-1)^l$. 
Thus, claim 3 gives
\be \langle l'\| \mathcal{Q} \| l\rangle = \sqrt{\frac{l+1}{2l+3}}\ \delta_{l',l+1} 
+ \sqrt{\frac{l}{2l-1}}\, \delta_{l',l-1} . \label{eq:97}
\ee

For $N=5$, the Weyl dimension formula gives
\be d(v) = \frac{1}{6} (v+1)(v+2) (2v+3) \ee
and, with the phase convention $e^{\mathrm{i}\phi(v)}=1$ (used in Ref.\
\cite{RTR}),  claim 3 gives
\be \langle v'\| \mathcal{Q} \| v\rangle = \sqrt{\frac{v+1}{2v+5}}\ \delta_{v',v+1}
+ \sqrt{\frac{v+2}{2v+1}}\ \delta_{v',v-1} \,. \label{eq:99}
\ee

\section{Combined matrix elements}\label{sect:combinedMEs}

Reduced matrix elements of $x = r\mathcal{Q}$ are obtained by combining the 
expression for $r$, given in terms of radial raising and lowering operators by equations
(\ref{eq:2r1}) and (\ref{eq:2r2}), with those for the matrix elements of
$\mathcal{Q}$, given by claim 2.
In particular, in the harmonic oscillator basis,
for which $\lambda_v = v+N/2$ (cf.\ fig.\ \ref{fig2a}), one obtains:
\bea &\langle \lambda_{v+1} \nu; v+1 \| x \| \lambda_v\nu;v\rangle
=\textstyle\half \langle \lambda_{v+1}\nu |A^\dag (\lambda_v-\half)
|\lambda_v,\nu\rangle \, \langle v+1\| \mathcal{Q}\| v\rangle , \label{eq:x1}\\
&\langle \lambda_{v+1}, \nu-1; v+1 \| x \| \lambda_v\nu;v\rangle
= \textstyle\half \langle \lambda_{v+1}\nu |A(-\lambda_v+\half)
|\lambda_v\nu\rangle \, \langle v+1\| \mathcal{Q}\| v\rangle ,&\\
&\langle \lambda_{v-1}, \nu+1; v-1 \| x \| \lambda_v\nu;v\rangle
=\textstyle\half \langle \lambda_{v-1},\nu+1 |A^\dag (-\lambda_v+\frac{3}{2})
|\lambda_v,\nu\rangle \, \langle v-1\| \mathcal{Q}\| v\rangle ,\\
&\langle \lambda_{v-1} \nu; v-1 \| x \| \lambda_v\nu;v\rangle
=\textstyle\half \langle \lambda_{v-1}\nu |A(\lambda_v-\frac{3}{2})
|\lambda_v,\nu\rangle \, \langle v-1\| \mathcal{Q}\| v\rangle .\label{eq:x4}
\eea
These identities have an immediate generalization.

\medskip {\bf Claim 4:} If $\alpha$ and $\beta$ index any radial wave functions,
the SO$(N)$ reduced matrix elements of the harmonic oscillator raising and lowering
operators between states of SO$(N)$ angular momentum $v$ and $v'$ are given by
\bea \langle \alpha v' \| c^\dag \| \beta v\rangle
&=& \textstyle\frac{1}{\sqrt{2}} \left[ \langle \alpha|A^\dag
(v+\half N-\half) |\beta\rangle \, \delta_{v',v+1}
+ \langle \alpha | A^\dag(-v-\half N+\frac{3}{2}) | \beta\rangle\,
\delta_{v',v-1}\right]   \nonumber
\\ && \times \langle v'\| \mathcal{Q}\| v\rangle  , \label{eq:cdagME}\\
\langle \alpha v' \| c \| \beta v\rangle
&=& \textstyle\frac{1}{\sqrt{2}} \left[ \langle \alpha|A
(-v-\half N+\half) |\beta\rangle \, \delta_{v',v+1}
+ \langle \alpha | A(v+\half N-\frac{3}{2}) | \beta\rangle\,
\delta_{v',v-1}\right]   \nonumber
\\ && \times \langle v'\| \mathcal{Q}\| v\rangle   .\label{eq:cME}
\eea
\medskip

With the substitution $x_i = \frac{1}{\sqrt{2}} (c_i^\dag + c_i)$, these results
follow immediately from equations (\ref{eq:x1}--\ref{eq:x4}) when the radial wave
functions are those of the harmonic oscillator basis. However, because  the radial
wave functions
$\{
\mathcal{R}^\lambda_\nu ; \nu = 0,1,2,\dots\}$ span the space of radial wave
functions for any value of $\lambda$, and because equations (\ref{eq:cdagME}) and
(\ref{eq:cME}) with
$|\beta v\rangle = |\lambda_v \nu; v\rangle$ and $|\alpha v'\rangle = |\lambda_{v'}
\nu'; v'\rangle$ hold for all values of $\nu$ and $\nu'$ they also hold for any
radial wave functions.

Because of the way the results of claim 4 are derived, they clearly incorporate the
identity
\be \langle \alpha v'\| x \|\beta v\rangle = 
\langle \alpha |r |\beta\rangle  \, \langle v' \| \mathcal{Q}\|v\rangle .\ee
What is more significant is that they also give the reduced matrix elements
\bea  
&\displaystyle\langle\alpha, v+1||| \hat p |||\beta v\rangle = -{\rm i}\hbar\, 
\langle \alpha \big|\Big[\frac{d}{dr} - 
\frac{v+\half N -\half }{r}\Big] \big|\beta\rangle  \, \langle v+1
\|\mathcal{Q}\|v\rangle
\,, & \label{eq:5.2Claim3}\\ 
&\displaystyle\langle \alpha, v-1||| \hat p |||\beta v\rangle = -{\rm i}\hbar\, 
\langle \alpha\big|\Big[\frac{d}{dr} + 
\frac{v+\half N -\frac{3}{2}}{r}\Big] \big|\lambda \nu\rangle  \,
\langle v-1 \|\mathcal{Q}\|v\rangle \,, &
\label{eq:5.2Claim4}
\eea
for the momentum operators
\be \hat p_i = -\mathrm{i}\hbar \frac{\partial}{\partial x_i} =
-\frac{\mathrm{i\hbar}}{\sqrt{2}}\, (c_i - c_i^\dag) .\ee

\section{Application to central force problems} 
\label{sect:application}

The characteristic property of a central force problem is that its Hamiltonian is
SO$(N)$-invariant.
For example, for a diatomic molecule the relevant Hilbert space is the
space of $\mathcal{L}^2(\Rb^3)$ wave functions in the relative coordinates of two
atoms. For a free molecule in an isotropic space, the Hamiltonian is
invariant under SO(3) rotations.
Thus, its Schr\"odinger equation reduces to an equation in a single radial variable.
Moreover, as already noted, the Hilbert spaces of radial wave functions are
independent of $N$.
Thus, methods developed for the solution of central force problems in $\Rb^3$ apply
more generally.

\begin{figure} [htp]
\epsfxsize=3.3in {\epsfbox{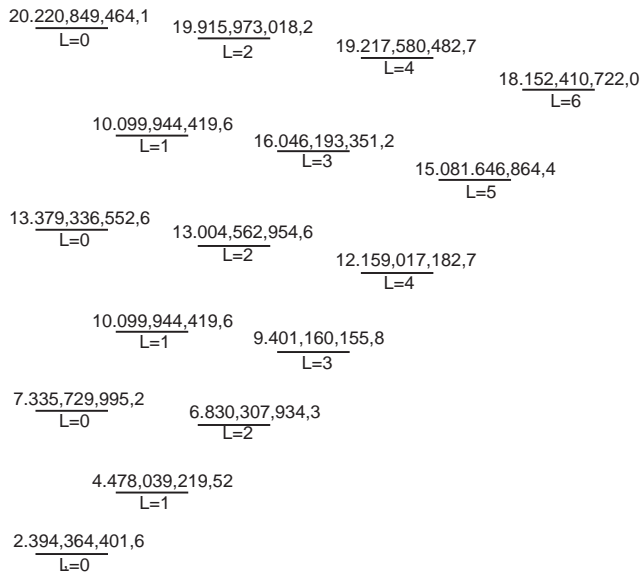}} 
  \caption{Low energy levels of the quartic oscillator in three dimensions.
\label{fig:3a}}  
\end{figure}

Consider, for example, the Hamiltonian of the quartic oscillator
\be \hat H =  -\frac{1}{2} \nabla^2 + r^4 . \ee 
Its spectrum was derived by Bell et al.\ \cite{BDW}  in two- and three-dimensional
spaces by diagonalization in a basis of eigenstates of the harmonic oscillator
Hamiltonian 
\be \hat H_{\mathrm{HO}} =  -\frac{1}{2a^2} \nabla^2 + \frac{1}{2}a^2r^2
\ee 
with $a=1$.
It has been considered more recently in five-dimensional space \cite{Arias}.
The addition of quartic terms, which lie in the
SU(1,1) enveloping algebra,  to a nuclear collective model Hamiltonian has also been
considered by several authors, e.g., \cite{VZ}.
Some  low-lying energy levels of the quartic oscillator in three dimensions
are shown  in Fig.\ \ref{fig:3a}. 
In repeating the calculations of ref.\ \cite{BDW} for all $L\leq 6$ energy levels
below 250 (in oscillator units), it was found that 21 s of computer time were
needed for each $L$ to compute these energy levels to an accuracy of 1 part in
$10^{12}$ when $a=1$ but only 2.6 s were needed with $a=1.6$.

A much larger gain is realized for the low-energy states of potentials with a minimum
at a non-zero value of
$r$ as the following application to the study of a phase transition in the nuclear
collective model shows.
Such a calculation was performed originally \cite{TR} in a five-dimensional
 harmonic oscillator basis.
However, subsequent studies \cite{RT} showed that it can be carried out much
more efficiently using modified oscillator bases and the algebraic matrix
elements associated with them.  This gain in efficiency is particularly relevant, as
discussed below, because it makes it practically possible to execute more
sophisticated calculations in which the radial and orbital degrees of freedom are
coupled.

The Hilbert space for the nuclear collective model is
$\mathcal{L}^2(\Rb^5)$ and the Hamiltonian  used in Ref.\ \cite{TR} was 
\be \hat H(\alpha) = -\frac{1}{2M} \nabla^2 +
\half M\left[ (1-2\alpha)r^2 +\alpha 
r^4\right] , \label{eq:V}\ee
where $r$ is a radial coordinate for $\Rb^5$ and $M$ a 
mass parameter. This Hamiltonian is
interesting because, as $\alpha$ passes through the critical value of 0.5, a phase
transition occurs from a spherical vibrational phase, corresponding to a minimum
value of
$V(r)$ at $r=0$, to a rotational-vibrational phase for $\alpha >0.5$, corresponding
to a minimum value of $V(r)$ at $\sqrt{(2\alpha- 1)/(2\alpha)}$ (in oscillator units
proportional to $\sqrt{M}$).

\subsection{Variational calculations}

\begin{figure} [htp]
\epsfxsize=4.3in {\epsfbox{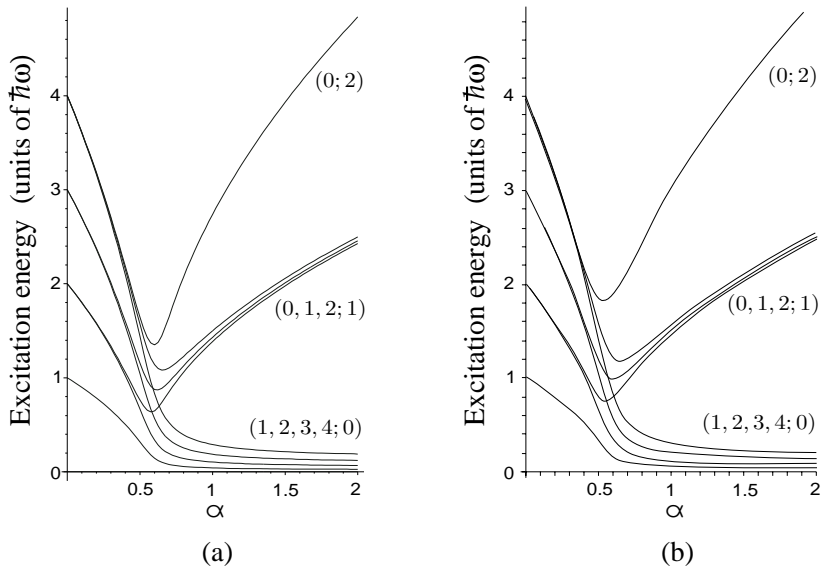}} 
  \caption{Comparison of the low-lying energy levels for the Hamiltonian $\hat
H(\alpha)$ with $M=100$ as computed (a) by diagonalization and (b) by taking
expectation values $\langle nvm |\hat H(\alpha)| nvm\rangle$ in single basis
states with variationally chosen parameters as described in the text. 
The numbers shown on the right hand side of each figure are the values of the
quantum numbers $(v_1,v_2,\cdots ; \nu)$ for each of the levels plotted. 
More detailed information is given for $\alpha = 1.5$ in figure \ref{fig:alpha1.5}. 
(Figure (a) was computed by P.S.\ Turner \cite{TR}.) 
\label{fig:3}} 
\end{figure}

The benefits from the new develoments are maximized by selecting basis wave functions
that are as close as possible to the eigenfunctions. For each value of $\alpha$ in
the Hamiltonian, an optimal orthonormal basis of states
$|n vm\rangle$ with wave functions of the form
\be\Phi_{ n vm}(r,\theta) = \frac{\sqrt{a_v}}{r^2}\, \mathcal{R}^{\lambda_v}_n
(a_vr)\, \mathcal{Y}_{vm}(\theta) ,\ee
is obtained by varying the parameters $a_v$ and $\lambda_v$ for each value of the
SO$(N)$ angular momentum $v$ to minimize the energy expectation values of
$\langle  0 vm|\hat H(\alpha) | 0 vm\rangle$.
The energy-levels obtained from the expectation values
\be E_{nvm}(\alpha) \approx \langle nvm |\hat H(\alpha) |nvm\rangle, \ee
with the $v$-dependent variationally-determined values of $a_v$ and $\lambda_v$, are
shown as functions of $\alpha$ in comparison with accurately computed values in 
Fig.\ \ref{fig:3}. 
The close correspondence between the results and those computed numerically by
diagonalization is remarkable.
Even in the transition region, where restriction to single basis states is least
successful, it does surprisingly well.  

\begin{figure} [htp]
\epsfxsize=3.8in {\epsfbox{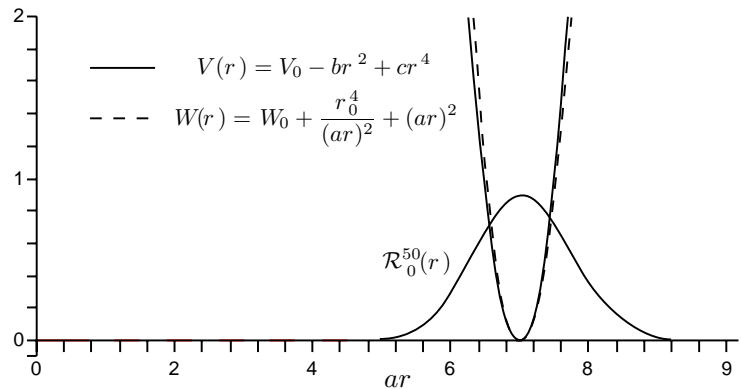}}
  \caption{Comparison of the potentials $V(r)$ and $W(r)$, as defined in the
text, for fitted values of $\lambda$ and $a$.
 $\{ \mathcal{R}^{50}_0\}$ is the radial wave function for the ground state of the
Hamiltonian with potential $W(r)$.
\label{fig:4}}       
\end{figure}

Why the variational calculations with single basis wave functions are so good is
illustrated in Fig.\ \ref{fig:4} which shows that a potential of the form
\be  V(r) =- br^2 + cr^4 ,\ee
which has a minimum at $r=r_0$,
is well fitted in the neighbourhood of its minimum by a potential
\be  W(r) = W_0 + \frac{r_0^4}{(ar)^2} + (ar)^2 ,\ee
with  a scale parameter  $a$ chosen to give $W(r)$ the curvature of the potential
$V(r)$ at $r_0$.
 The radial eigenfunction for the ground state of a Hamiltonian with potential
$W(r)$ is given by $\mathcal{R}^{50}_0(r)$ (in suitable units of $r$)
and found to be an excellent approximation to that for the potential
$V(r)$.  
  
\begin{figure} [htp]
\epsfxsize=2.7in {\epsfbox{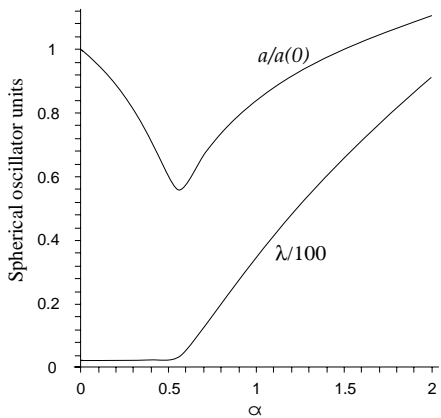}} 
  \caption{Values of $\lambda$ and the inverse width parameter $a/a(0)$, where
$a(0)=\sqrt{M}$, for the variational ground-state wave functions used in Fig.\
\protect\ref{fig:3} (b). 
\label{fig:5}}  
\end{figure}

The values of $a$ and $\lambda$ which minimize the 
variational ground-state energy are shown in Fig.\ \ref{fig:5}.
The figure shows that the wave function is essentially
that of a spherical vibrator for $\alpha < 0.5$ but that its width increases with
$\alpha$ until the point at which the curvature of the potential 
$V(r) =\half M\left[ (1-2\alpha)r^2 +\alpha 
r^4\right]$ at its $r=0$ minimum vanishes.  
With further increase in $\alpha$, the wave function becomes that of a
rotor-vibrator with equilibrium deformation given by  
$r_0 =[(\lambda-1)^2-9/4]^{1/4}/a$

\subsection{Results from numerical diagonalization}

With the analytical expressions given in Sect.\ \ref{sect:SU11}, variational
calculations of the type shown above are easy to implement and give insightful
information about the states of the system.
Moreover, they are readily extended to obtain more precise results by
diagonalization of the Hamiltonian in an optimal basis.

 If the objective is to obtain the energy
eigenstates for the $n$ lowest-energy states as accurately as possible when computed
in a space spanned by the first $N$ states of an ordered basis, then a simple
prescription  for  selecting  a near to optimal  basis among a set of possible
choices  is to select the basis which minimizes the computed energy for the highest
among the $n$ lowest energy states.
\medskip

Other prescriptions can obviously be defined such as, for example,
minimizing the sum of the energies of the lowest $n$ states. However, it is not clear
that much can be gained over the above simple prescription.

\begin{figure} [htp]
\epsfxsize=5in {\epsfbox{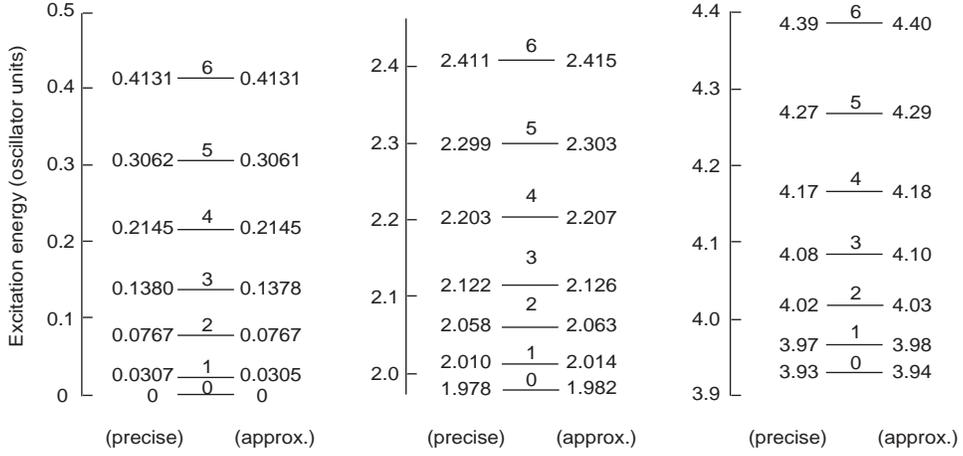}}  
  \caption{Energy-level spectrum of the Hamiltonian $\hat H(\alpha)$ for $\alpha =
1.5$ and $M=100$ for the three lowest-energy states of SO(5) angular momentum $v\leq
6$. The precise energies were computed by diagonalization with 100 basis states. 
The approximate values were computed with just 5 basis states chosen as described in
the text.
\label{fig:alpha1.5}}  
\end{figure}

For $\alpha\gtrsim 1.5$, for which the model nucleus described by the Hamiltonian
$\hat H(\alpha)$ of equation (\ref{eq:V}) is clearly non-spherical for $M=100$, it was
found that the energy of the lowest energy state of each SO(5) angular momentum
$v$ was obtained to within 1\% accuracy with just one modified oscillator basis wave
function. To achieve this level of accuracy with a conventional
harmonic oscillator basis requires $\gtrsim 22$  harmonic oscillator basis wave
functions (depending on the value of $\alpha$). 
Fig.\ \ref{fig:alpha1.5} shows the excitation energies of the three lowest-energy
states for each  $v\leq 6$ computed precisely and with just 5 basis states.

The basis states were chosen for the $v=0$ states according to the above given
prescription.
This prescription could have been followed for each value of $v$.
However, in practice, one will want to be able to compute electromagnetic transition
matrix elements.  
Having fixed $\lambda$ for the $v=0$ states (which turned out to be 57) we then chose
 $\lambda=57$ for all even $v$  and $\lambda = 58$ for all odd $v$ for the reasons
discussed in  Sect.\ \ref{sect:parity}.

To obtain parallel results to the same level of accuracy in a harmonic oscillator
basis requires $\sim 25$ basis states and a computation that takes $\sim 10$ times as
long. For practical purposes, a reduction in the required number of radial basis
states is particularly advantageous in treating systems in which there is a coupling
between the orbital and radial degrees of freedom; this is because the product of
the needed number of radial and orbital basis states can reach a large number.

\section{More general applications} 

\subsection{An $N=3$ example}

In  many problems of interest, the SO$(N)$ rotational invariance is broken.
For example, in $\Rb^3$, it is of interest to study  splittings
of molecular energy levels by a crystal field.
This might be done with a Hamiltonian of the type
\be \hat H = -\frac{1}{2M} \nabla^2 +
V(r) + \chi r^2(3\cos^2\theta - 1). \label{eq:MolH}\ee

With the $M=0$ component of the $L=1$ spherical tensor $\mathcal{Q}$ given,
according to equation (\ref{eq:Q1}), by $\mathcal{Q}_{10} = \cos\theta$, the function
$3\cos^2\theta -1$ is proportional to the $M=0$ component of the
$L=2$ tensor $(\mathcal{Q}\otimes\mathcal{Q})_2$, i.e.,
\be 3\cos^2\theta -1 = \sqrt{6}\, (\mathcal{Q}\otimes\mathcal{Q})_{20} .\ee
The reduced matrix elements of the tensor $(\mathcal{Q}\otimes\mathcal{Q})_2$,
obtained from those of $\mathcal{Q}$ by  Racah recoupling, are given by
\be \langle l_2 \| (\mathcal{Q}\otimes\mathcal{Q})_2 \| l_1\rangle =
\sum_{l} \sqrt{5(2l+1)}\, W(l_1 2,l_2 2; l\,2)\, \langle l_2 \|\mathcal{Q}\| l\rangle
\,\langle l \|\mathcal{Q}\| l_1\rangle .
\ee
It follows that 
\be \langle l_2m |(3\cos^2\theta -1) | l_1m\rangle =
\sum_{l} \sqrt{30(2l+1)}\, (l_1 m,20|l_1m) W(l_1 2,l_2 2; l\,2) \langle l_2
\|\mathcal{Q}\| l\rangle
\langle l \|\mathcal{Q}\| l_1\rangle,
\ee
where $(l_1 m,20|l_1m)$ is an SO(3) Clebsch-Gordan coefficient.
These matrix elements are readily evaluated with the $\langle l'\|
\mathcal{Q}\|l\rangle$ matrix elements given by equation (\ref{eq:97}).

The effect on a spherical molecule of putting it into a quadrupole field is to
deform it somewhat.  However, if the molecule is already deformed the primary effect
 is to align the deformation of the molecule with  the field.
Thus, for a well-deformed diatomic molecule such as the HCl molecule, an informative
way to study the spectrum and eigenfunctions of the Hamiltonian (\ref{eq:MolH}) would
be to start by solving for the Hamiltonian
 \be \hat H_0 =  -\frac{1}{2M} \nabla^2 +
V(r) + \chi r_0^2(3\cos^2\theta - 1) . \ee
This will give the splittings of energy levels due to the alignment effect.
Solutions for the perturbed Hamiltonian
\be \hat H = \hat H_0 + \chi (r^2-r_0^2)(3\cos^2\theta - 1) \ee
will then give the added perturbations of the spectra coming from the
rotation-radial vibration coupling interactions and provide information about the
rigidity of the molecule.

\subsection{An $N=5$ example} 

For $N>3$, there may be terms in the Hamiltonian which break the SO$(N)$ invariance
but retain  rotational invariance with respect to a suitably defined $\mathrm{SO}(3)
\subset \mathrm{SO}(N)$ subgroup.
For example,  a Hamiltonian of interest in the nuclear collective model is of the
form
\be \hat H(\alpha) = -\frac{1}{2M} \nabla^2 +
\half M\left[ (1-2\alpha)r^2 +\alpha 
r^4\right] +\kappa r^3(\mathcal{Q}\otimes\mathcal{Q}\otimes\mathcal{Q})_{0} ,
\label{eq:122}\ee
where $(\mathcal{Q}\otimes\mathcal{Q}\otimes\mathcal{Q})_{0}$ is triple
product of $\mathrm{SO}(5) \supset \mathrm{SO}(3)$, $v=1,L=2$, 
$\mathcal{Q}$ tensors coupled to SO(3) angular momentum $L=0$.
This SO(3) coupled product is the $L=0$ component of a $v=3$, SO(5) tensor.

The  matrix elements of
$(\mathcal{Q}\otimes\mathcal{Q}\otimes\mathcal{Q})_{0}$
are  given by
\bea    \langle v'\alpha'LM
|(\mathcal{Q}\otimes\mathcal{Q}\otimes\mathcal{Q})_{0}|v\alpha LM\rangle
&=&\raisebox{0.5ex}{
$\displaystyle\sum_{{v_1\alpha_1L_1} \atop {v_2\alpha_2L_2}}$ }
(-1)^{L+L_2}\sqrt{\frac{(2L_1+1)(2L_2+1)}{2L+1}}
 W(L2L_2L;L_12) \nonumber\\
&&\times \langle v'\alpha'L\|{\cal Q} \| v_2\alpha_2L_2\rangle
\langle v_2\alpha_2L_2\|{\cal Q} \| v_1\alpha_1L_1\rangle
\langle v_1\alpha_1L_1\|{\cal Q} \| v\alpha L\rangle .
\eea
Note, however, that the reduced matrix elements appearing in this expression are 
SO(3)-reduced matrix elements; they are related to the SO(5)-reduced matrix
elements of equation (\ref{eq:99}) by the expression
\be
\langle v_1\alpha_1L_1\|{\cal Q} \| v\alpha L\rangle = (v\alpha L, 12 \|
v_1\alpha_1L_1) \langle v_1\|{\cal Q} \| v\rangle ,\ee
where $(v\alpha L, 12 \| v_1\alpha_1L_1)$ is an SO(5) Clebsch-Gordan
coefficient in an SO(3) basis.
An algorithm for computing such CG coefficients and tables of values has been
given in Ref.\ \cite{RTR}.  

\begin{figure} [htp]
\epsfxsize=4in {\epsfbox{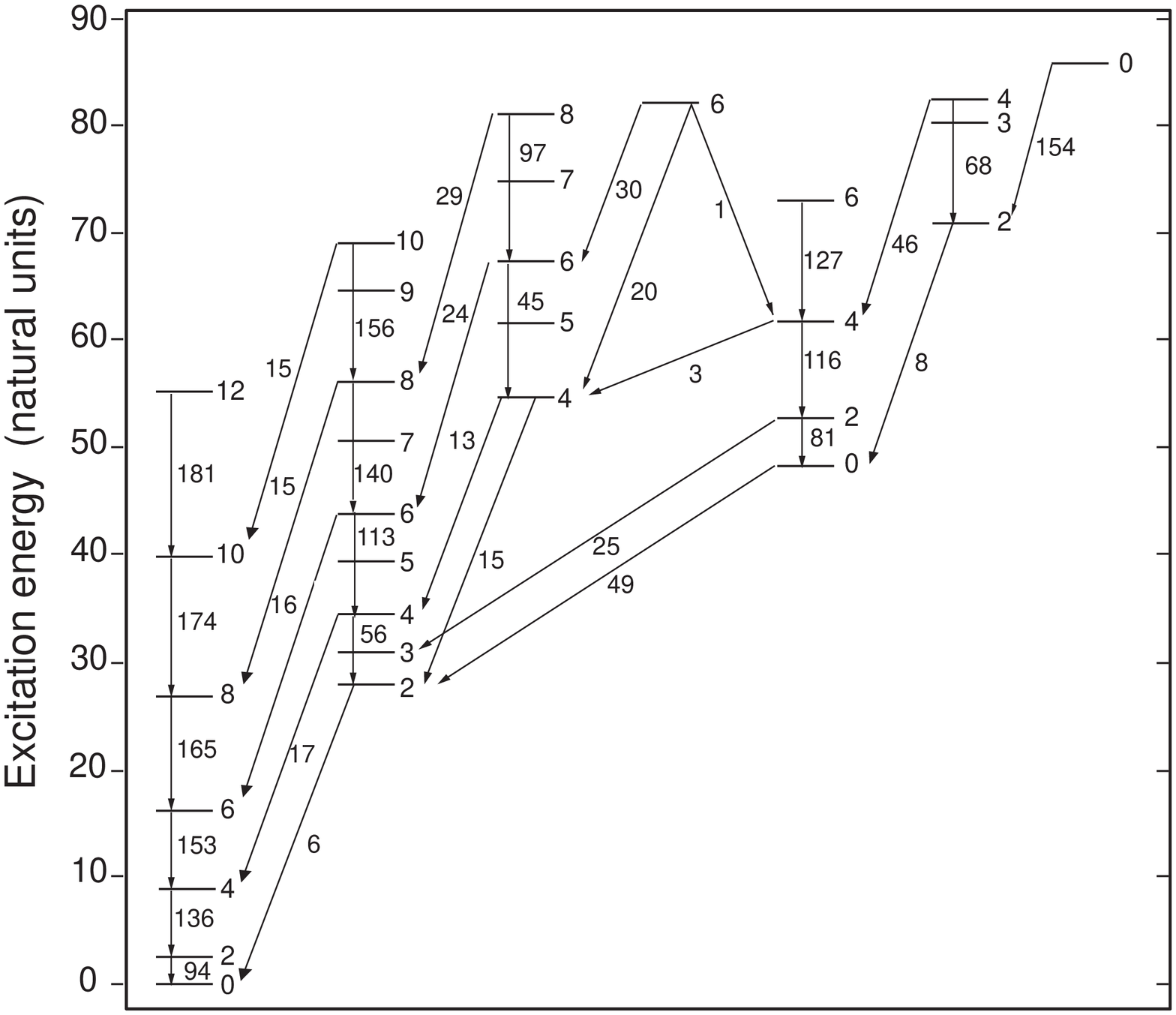}} 
  \caption{Low energy-level spectrum of the Hamiltonian of equation
(\protect\ref{eq:122}) with a large value of $\alpha$ and $\kappa = 50 \sqrt{2/35}$. 
Each energy level is labelled by its SO(3) angular momentum; because of the
interaction, the SO(5) angular momentum is no longer a good quantum number.
Transition rates for electric quadrupole gamma-ray transitions indicated by arrows
 are shown in natural units beside the arrows; these transition rates are known in
nuclear physics as {\em reduced E2 transition rates\/}. 
\label{fig:6}}
\end{figure}

The low-energy level spectrum of the Hamiltonian (\ref{eq:122}),  calculated
 with a large value of $\alpha$ \cite{Rowe}, is shown in Fig.\ \ref{fig:6}.       
The figure shows a ground-state rotational band and a sequence of excited (so-called
gamma-) vibrational bands of the type given by the phenomenological Bohr-Mottelson
nuclear collective model
\cite{BM}.            

\section{Concluding remarks}\label{sect:conc}

It has been shown that the matrix elements of polynomial Hamiltonians on a Euclidean
space $\Rb^N$ can be determined algebraically to within SO$(N)$ Clebsch-Gordan
coefficients.

The results obtained highlight the importance of algorithms for computing
CG coefficients for various SO$(N)$ groups.
Currently these coefficients are available for $N\leq 6$.
Explicit expressions for SO(3) coefficients were given already in the 1931
edition of Wigner's book on Group Theory \cite{Wig} and can now be found in
almost any book on angular momentum theory.
The coefficients for SO(4),  locally isomorphic to SO(3) $\times$ SO(3),
are readily obtained from combinations of the SO(3) coefficients \cite{BGW}.
An algorithm for calculating the needed coefficients for SO(5)  has recently
been given \cite{RTR} in an SO(3) basis defined by regarding the fundamental
5-dimensional $v=1$ irrep of SO(5) as carrying in irreducible $L=2$ irrep of SO(3).
This algorithm for SO(5) CG coefficients extends to SO(6) in an SO(5)
$\supset$ SO(3) basis by regarding the fundamental 6-dimensional irrep of SO(6) as
spanning a sum of
$v=0$ and $v=1$ SO(5) irreps and noting that the $v=0$ irrep of SO(5) is the trivial
identity irrep. CG coefficients could undoubtedly be used with advantage for higher
SO$(N)$ groups.
For example,  models for octupole vibrations and rotations of
nuclei or molecules with octupole deformations could be formulated on $\Rb^7$ and
would require SO(7) CG coefficients.

It will interesting to see if the techniques introduced have a parallel extension to
hydrogenic and modified hydrogenic systems in light of the recent developments of
Fortunato and Vitturi \cite{FV}.

\acknowledgements{The author is pleased to acknowledge helpful suggestions and
references from J.L.\ Wood and J.\ Karwowski.}

\end{document}